\documentclass[cits]{PoS}
\usepackage{amssymb,fontenc,times,mathptmx,graphicx}
\usepackage{amsmath,latexsym}
\usepackage{axodraw,mathrsfs}
\usepackage{bbm}
\usepackage{macros}

\title{%%% remove in PoS-version!!!
\vspace*{-2cm}
\begin{minipage}{\textwidth}
\begin{flushright}
{\normalfont{\small{DESY 08-136}}}\\
{\normalfont{\small{LTH 805}}}\\
\end{flushright}
\end{minipage}\\[35pt]
Wilson twisted mass fermions in the epsilon regime} 

\ShortTitle{Wilson twisted mass fermions in the epsilon regime}

\author{K. Jansen, A. Nube,\\
        DESY Zeuthen, Platanenallee 6 \\
        D-15738 Zeuthen, Germany\\
        E-mail: \email{karl.jansen@desy.de},\email{annube@ifh.de}}

\author{\speaker{A. Shindler}\\
        Theoretical Physics Division, \\
        Dept. of Mathematical Sciences, \\
        University of Liverpool \\
        Liverpool L69 7ZL, UK \\
        E-mail: \email{shindler@liv.ac.uk}}

\vspace*{-10mm}

\vspace*{7mm}

\abstract{
  \begin{center}
    \includegraphics[draft=false,scale=1]{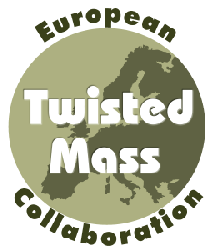}
  \end{center}
  \vskip 0.5cm
  In this proceeding contribution we report on the ongoing effort to understand 
  and simulate Wilson twisted mass fermions in the so called \mbox{$\epsilon$ regime}.
}

\FullConference{The XXVI International Symposium on Lattice Field Theory \\
         July 14 - 19, 2008\\
         Williamsburg, Virginia, USA}

\begin{document}

\section{Introduction}
\label{sec:intro}
Simulations in the $\epsilon$ regime are complementary
to standard large volume simulations.
They allow to extract low energy constants of the chiral Lagrangian,
in some cases with less contaminations from chiral logs coming
from higher order corrections.
For a long time it has been thought that simulations
in the $\epsilon$ regime are restricted to chirally invariant
lattice formulations.
In ref.~\cite{Jansen:2007rx} we have argued that actually this is not the case,
and that simulations in the $\epsilon$ regime can be performed
also with non chirally invariant lattice actions such as Wilson like fermions.

In particular in~\cite{Jansen:2007rx} we suggested that a combination 
of algorithmic and theoretical understanding of Wilson twisted mass
makes it possible to actually perform simulations in the $\epsilon$ regime
with Wilson twisted mass fermions.

Recently it has been shown that with suitable and related algorithmic ideas~\cite{Hasenfratz:2008fg}
it possible to reach or get close to the $\epsilon$ regime also with improved Wilson fermions.
At this lattice conference further results in this directions have been presented~\cite{lat08talks:2008wi}.

In this proceeding we consider a second lattice spacing and we extend to NLO
the analysis performed in~\cite{Jansen:2007rx}.
Our setup is a $L^3 \times T$ euclidean lattice with spacing $a$.
The lattice action
\be
S[\chi,\chibar,U] = S_G[U] + S_F[\chi,\chibar,U],
\ee
is the sum of the so called tree-level improved Symanzik gauge action~\cite{Weisz:1982zw}
\be
S_G[U] = \frac{\beta}{3}\sum_x\left\{b_0\sum_{\mu<\nu}
\mathbb{R}{\rm e}~ \Tr \left[ \mathbbm{1} - P^{(1\times 1)}(x;\mu,\nu)\right] + b_1 \sum_{\mu \neq \nu}
\mathbb{R}{\rm e}~ \Tr \left[ \mathbbm{1} - P^{(2\times 1)}(x;\mu,\nu)\right] \right\} ,
\ee
where
\be
b_0 = 1-8b_1, \quad b_1 = - {1 \over 12},
\ee
with Wilson twisted mass~\cite{Frezzotti:2000nk}
\be
  S_{\rm F}[\chi,\chibar,U] =a^4\sum_x\chibar(x)\Big[D_{\rm W} + i\mu_{\rm q}\gamma_5\tau^3\Big]\chi(x),
\label{eq:WtmQCD}
\ee
where
\be
D_{\rm W} = \frac{1}{2}\{\gamma_\mu(\nabla_\mu + \nabla^*_\mu) -a  \nabla^*_\mu\nabla_\mu\} + m_0.
\label{eq:Wilson}
\ee

The basic idea of~\cite{Jansen:2007rx} is that by sampling all topological sectors in the ensemble generation,
it is not necessary to have an unambiguous definition of topology at finite
lattice spacing. To achieve this goal it was suggested~\cite{Jansen:2007rx} to use 
a PHMC algorithm~\cite{Frezzotti:1997ym} treating the low modes exactly and reweighting the observables.
This could allow to perform simulations at very low quark masses without 
encountering instabilities or metastabilities. 
\section{$\epsilon$ expansion}
\label{sec:epsilon}
Lowering the quark mass at finite lattice spacing 
with Wilson-like fermions requires a detailed understanding
of the interplay between the genuine chiral behaviour induced
by the 'pion' dynamics and the one generated by cutoff effects.
A review on the phase diagram and cutoff effects with Wilson twisted mass (Wtm)
can be found in ref.~\cite{Shindler:2007vp}.
In the $\epsilon$ regime this is equivalent to saying that it is necessary 
to understand the coupling of the zero modes with the relevant operators
describing the effect of the lattice artifacts.
The actual values of the lattice spacing,
the physical volume and the quark mass 
determine the appropriate power counting, which ought to be
used to perform computations using chiral effective theories.
In the continuum the exact integration over the constant 
zero modes can be achieved in the chiral effective theory modifying the
$p$ regime power counting, in the so called $\epsilon$ expansion
where the would-be pion mass is small compared
to the linear size of the box
\be
 \frac{1}{T} = {\rm O}(\epsilon) , \quad \frac{1}{L} = {\rm O}(\epsilon) , \quad M_\pi = {\rm O}(\epsilon^2) .
\ee
As a result of the exact integration the order parameter, or the equivalent ratio
$R = \frac{\langle \bar{q}q\rangle}{B_0F^2}$, vanishes in the chiral 
limit at fixed finite volume~\cite{Gasser:1987ah}, obtaining
restoration of chiral symmetry.
One possible way to include the effects of the lattice artifacts in this analysis
is to include with an appropriate power counting the lattice spacing.
Here we modify the standard power counting in the following way~\cite{Andrea:200x}
\be
  M = O(\epsilon^4), \quad \frac{1}{L} = O(\epsilon), \quad
  \frac{1}{T} = O(\epsilon) \quad {a^2 = O(\epsilon^4)},
\ee
where $M$ indicates generically a quark mass.
The partition function at leading order reads
\be
  \mcZ = \int \mcD[U_0] {\rm e}^{\frac{c_1V}{2}\Tr\left[U_0 + U_0^\dagger\right] - 
  \frac{c_2V}{4}\Tr\left[U_0 + U_0^\dagger\right]^2+\frac{c_3V}{2}\Tr\left[i\tau^3\left(U_0^\dagger - U_0\right)\right]},
\ee
where the scaling variables are
\be
  z_1 = c_1V = B_0F^2m'V , \qquad z_2 = c_2V = -\frac{F^2w'Va^2}{4}, \qquad z_3 = c_3V = B_0F^2\mu_{\rm R} V . 
\ee
To argue that this is a proper power counting for actual numerical
simulations we list here some values
\be
  M \simeq 5 {\rm MeV}, \quad a \simeq 0.1 {\rm fm}, \quad L \simeq 1.5 {\rm fm} 
\ee
\be
  \quad  F\simeq 90 {\rm MeV}, \quad B_0 \simeq 5.5 {\rm GeV}, \quad |w'| \simeq (570 {\rm MeV})^4.
\ee
Using these values to estimate the relevant scaling variables in this regime
one obtains
\be
  M F^2 B_0 V \simeq 0.75, \quad a^2F^2|w'|V \simeq 0.75, \quad \frac{MB_0}{a^2|w'|} \simeq 1,
\ee
which indicates that this is an appropriate power 
counting.\footnote{If the lattice spacing is much smaller a different
power counting ought to be used where the lattice artifacts only appear
at NNLO.}The chiral condensate can be computed in the standard way
%\be
%  R = \frac{\langle \bar{q}q \rangle}{B_0F^2}
%\ee
\be
  R = \frac{1}{N_{\rm f}}\frac{\partial}{\partial z_3}\log\mcZ, \qquad z_1=0,
\ee
\begin{figure}[h]
\vspace{-0.8cm}
  \begin{center}
    {\hspace{-1.0cm}\includegraphics[height=2.0cm,angle=270,scale=4]{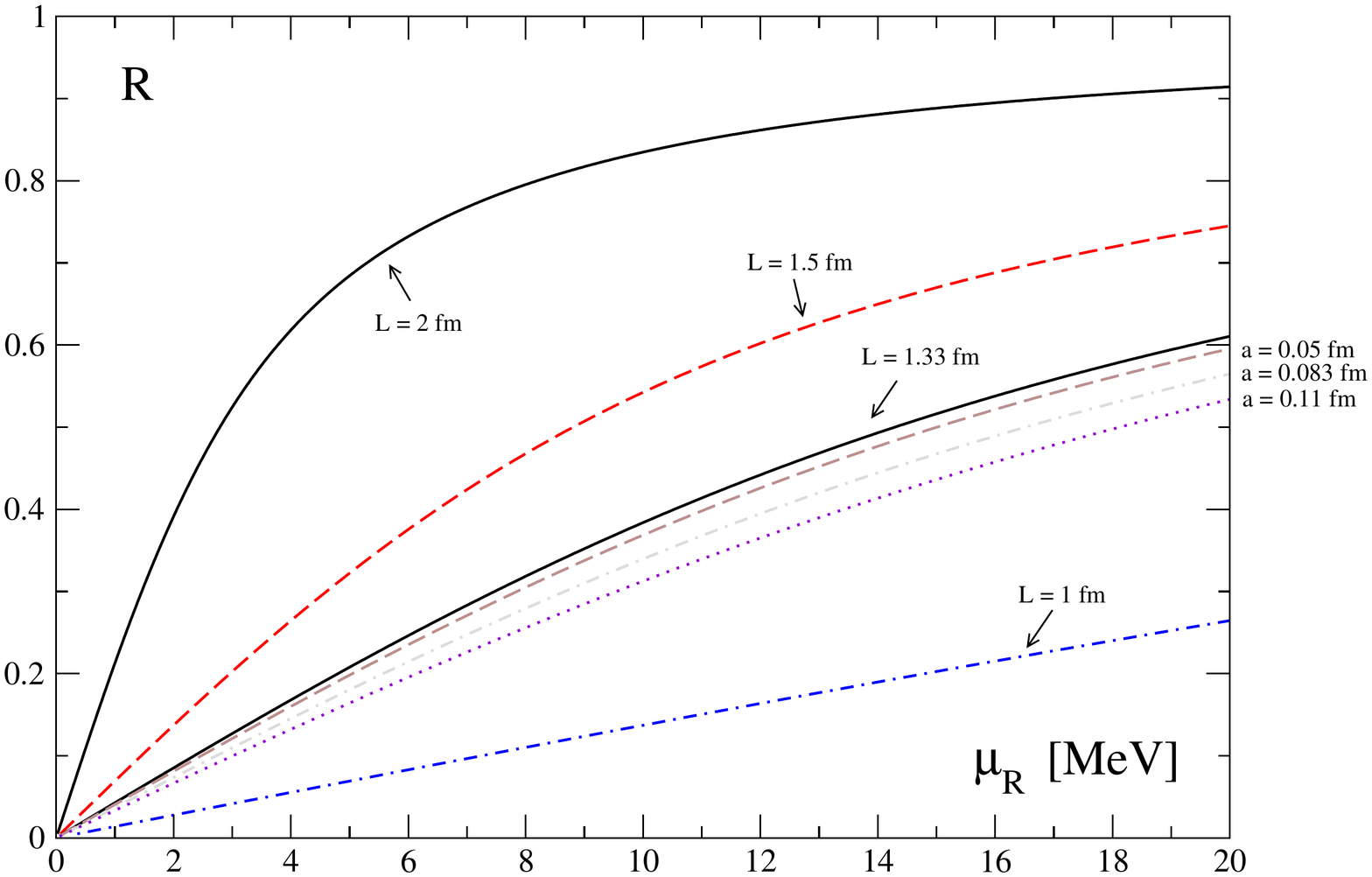}
    \includegraphics[height=2.0cm,angle=270,scale=4]{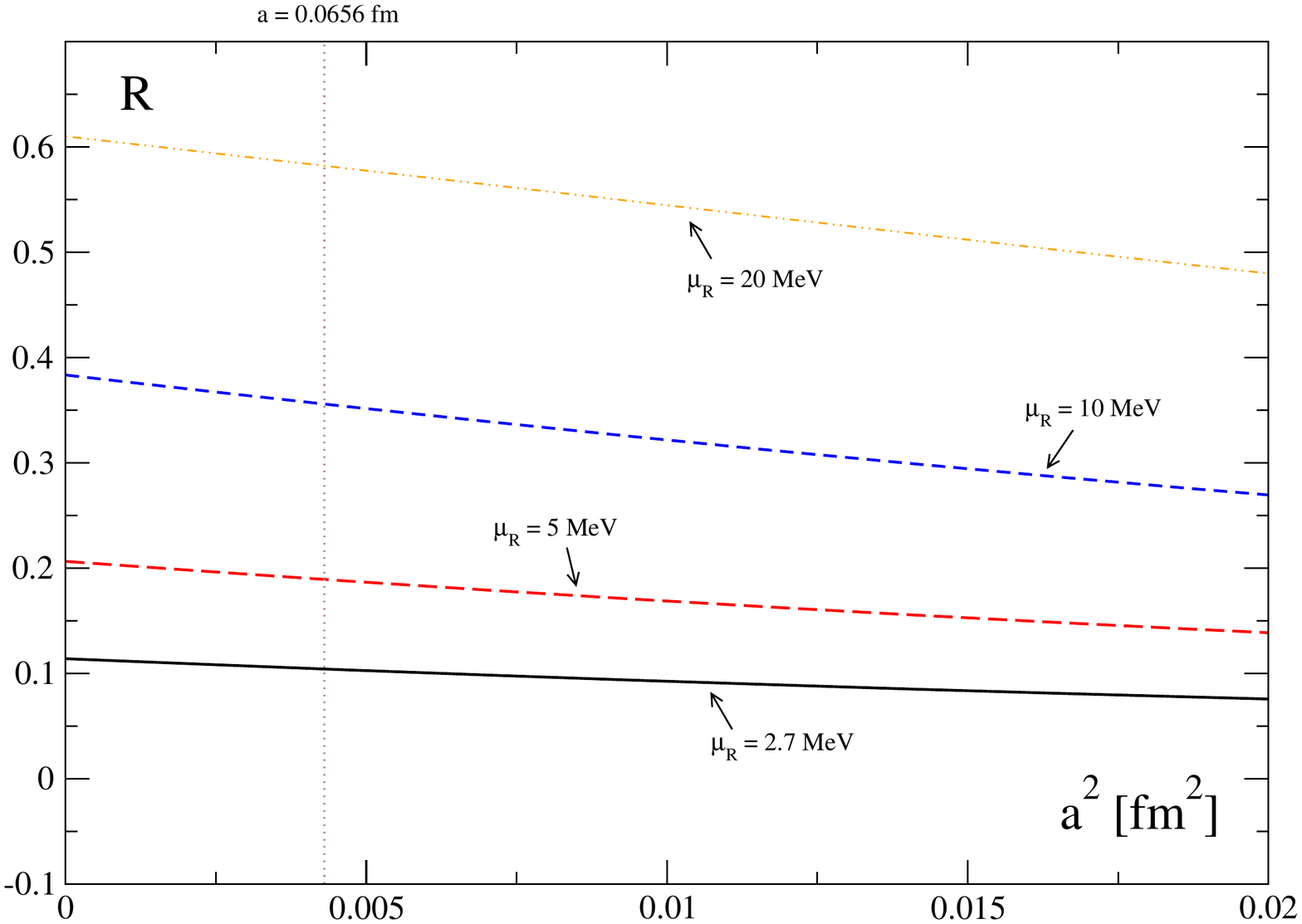}}

    \vspace{-0.8cm}
    \caption{Quark mass (left plot) and lattice spacing (right plot) dependence
      for the single flavour chiral condensate normalized with its LO value 
      in the continuum and infinite volume}
    \label{fig:cond_eps}
  \end{center}
\end{figure}
and fig.~\ref{fig:cond_eps} shows the quark mass (left plot) and lattice spacing
(right plot) dependence of the chiral condensate. 
We can certainly conclude that the dependence on the quark mass
is, as expected, smooth, and the cutoff effects are under control.
Extension of this computation to NLO including standard 2-point functions
is currently in progress~\cite{Andrea:200x}.
The power counting introduced is general and valid also for plain Wilson
fermions ($z_3=0$). The same power counting could be used to develop an expansion
with staggered fermions and to check the chiral properties
of the spectrum in the presence of roots of the staggered determinant.
%    \begin{itemize}
%    \item<1-> General power counting: valid also for Wilson fermions
%    \item<1-> extension to NLO in progress
%    \item<1-> No phase transition/minimal twisted mass in the $\epsilon$ regime
%    \item<1-> It could be used to attack other problems like: interplay between $a$ and $V$ in the eigenvalues
%      distribution
%    \item<1-> Alternative way to extract the LEC $|w'|$ which parametrized the O($a^2$) effects
%    \item<1-> Is this the correct power counting?
      %      \item<1-> Simulations in the $\epsilon$ regime allow extraction of LEC
      %      with kinematically suppressed NNLO
      %    \item<1-> Which contribution at NLO?
      %    \item<1-> N(N)LO $\rightarrow$ uncertainties on the PCAC mass 
%    \end{itemize}
\section{Numerical results}
\label{sec:numerics}
\begin{figure}[h]
\vspace{-0.8cm}
  \begin{center}
    {\hspace{-1.0cm}\includegraphics[height=2.0cm,angle=270,scale=4]{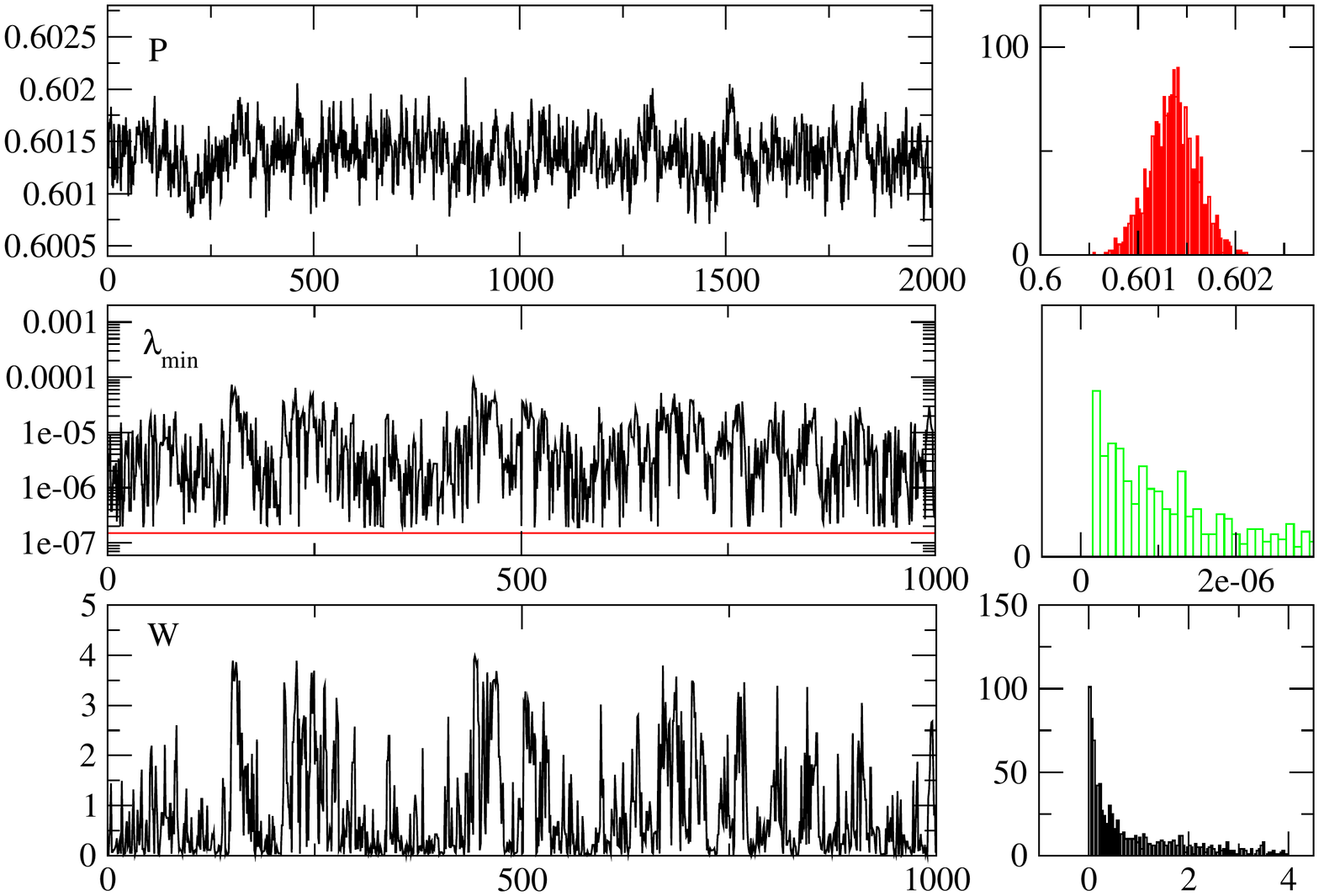}
%    \vspace{-1.5cm}
    \includegraphics[height=2.0cm,angle=270,scale=4]{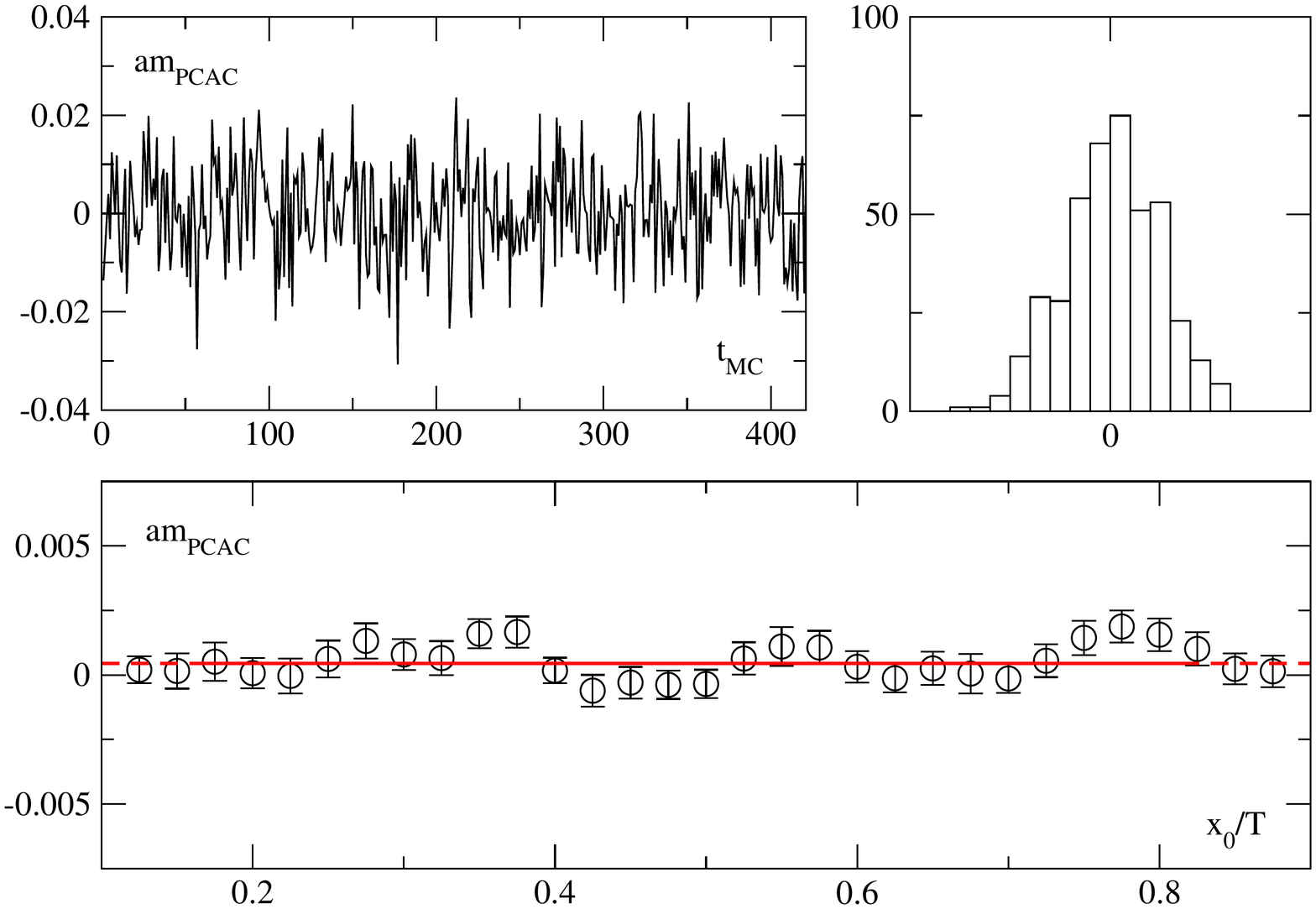}}
    \vspace{-0.8cm}
    \caption{{\it Left plot}: MC histories and distributions of the plaquette (first strip) smallest eigenvalue
      (second strip) and reweighting factor (third strip). The smallest eigenvalue is compared with
      the infrared cutoff provided by the twisted mass (horizontal line).
      {\it Right plot}: MC history and distribution at $x_0 = T/4$ (first strip)
      together with the euclidean time dependence of the PCAC mass (second strip).}
    \label{fig:algo_pcac}
  \end{center}
    \vspace{-0.5cm}
\end{figure}

%The algorithm used is a PHMC with exact reweighting. 
%A generic correlation function
%\be
%  \langle {\mathcal O} \rangle = \frac{1}{{\mathcal Z}}\int {\mathcal D}[U] {\rm
%  e}^{-S_G[U]} \det\left(QQ^\dagger \left[U\right]\right) {\mathcal O}\left[U\right] \qquad 
%  Q = \gamma_5\left[D_{\rm W} + i\mu_{\rm q} \gamma_5\right],
%\ee
%with $Q$ single flavour operator, can be computed with a reweighted determinant
%\be
%  \det\left(QQ^\dagger \left[U\right]\right)  = \frac{\det\left[QQ^\dagger 
%  P_{n,\tilde{\epsilon}}\left(QQ^\dagger \right)\right]}{\det\left[P_{n,\tilde{\epsilon}}\left(QQ^\dagger \right)\right]} , 
%  \quad P_{n,\tilde{\epsilon}}\left(QQ^\dagger \right) \simeq \left[QQ^\dagger\right]^{-1}.
%\ee
%Here
%\be
%  P_{n,\tilde{\epsilon}}\left(QQ^\dagger \right) \simeq \left[QQ^\dagger\right]^{-1} 
%  \qquad \left\{\lambda\right\} \in[\tilde{\epsilon},1],
%\ee
%is a polyinomial approximation of the $\left[QQ^\dagger\right]^{-1}$ normalized to have
%the largest eigenvalue equal to 1.
%Observables can then be computed with an exact reweighting factor
%\be
%  \langle {\mathcal O} \rangle = \frac{\langle {\mathcal O} W \rangle_P}{\langle
%  W \rangle_P}, \quad W =
%  \det\left[QQ^\dagger P_{n,\tilde{\epsilon}}\left(QQ^\dagger \right)\right] \simeq
%  \prod_{\lambda_i < \tilde{\epsilon}}\left[\lambda_iP_{n,\tilde{\epsilon}}\left(\lambda_i\right)\right] ,
%\ee
%which containd the smallest O($20$) eignevalues.
%This procedure allows a better sampling of the configuration space,
%and removes all the issues with possible instabilitites/metastabilities~\cite{lat07}.
Details of the algorithm used to generate the gauge ensemble can be
found in ref.~\cite{Jansen:2007rx}.
In this proceeding we complement the results obtained in~\cite{Jansen:2007rx}
with a second lattice spacing with a NLO analysis.
The inversions for the quark propagator have been performed with a stochastic
$Z_2 \times Z_2$ source located randomly along the euclidean time.
Table~\ref{tab:setup} summarizes the simulation setup.
\begin{table}
\begin{center}
 \begin{tabular}{|l|l|l|l|l|}\hline
   \multicolumn{1}{|c|}{\ $\beta$ \; \ } & 
   \multicolumn{1}{|c|}{\ $\kappa$ \; \ } & 
   \multicolumn{1}{|c|}{\ $L/a$ \; \ } & 
   \multicolumn{1}{|c|}{\ $T/a$ \; \ } & 
   \multicolumn{1}{|c|}{\ $a\mu_{\rm q}$ \; \ } 
   \\ 
   \hline
   \ $4.05$   & \   $0.157010$   & \  $20$  & \  $40$  & \  $0.00039$   \\
   \hline
 \end{tabular}
\end{center}
%\newline
\vspace{-0.5cm}
\begin{center}
 \begin{tabular}{|l|l|l|l|}\hline
   \multicolumn{1}{|c|}{\ $N_{\rm traj}$ \; \ } & 
   \multicolumn{1}{|c|}{\ $N_{\rm ana}$ \; \ } & 
   \multicolumn{1}{|c|}{\ $\tau_{\rm int}(P)$ \; \ } & 
   \multicolumn{1}{|c|}{\ $\tau_{\rm int}(m_{\rm PCAC})$ \; \ }
   \\ 
   \hline
   \ $2500$   & \   $421$   & \  $\sim 0.5$  & \  $\sim 0.5$  \\
   \hline
 \end{tabular}
\end{center}
% \newline
\vspace{-0.5cm}        
\begin{center}
 \begin{tabular}{|l|l|l|l|}\hline
   \multicolumn{1}{|c|}{\ $r_0/a$ \; \ } & 
   \multicolumn{1}{|c|}{\ $a [{\rm fm}]$ \; \ } & 
   \multicolumn{1}{|c|}{\ $L [{\rm fm}]$ \; \ } & 
   \multicolumn{1}{|c|}{\ $am_{\rm PCAC} $ \; \ } 
   \\ 
   \hline
   \ $6.61(3)$   & \   $0.0656(11)$   & \  $1.31$  & \  $0.00045(12)$  \\
   \hline
 \end{tabular}
\end{center}
\vspace{-0.5cm}
\caption{Summary of the simulation setup and of the basic ensemble parameters.}
 \label{tab:setup}
\end{table}
In the left plot of fig.~\ref{fig:algo_pcac}, we show in the first strip the plaquette MC history and its
distribution. In the second strip we show the MC history and distribution
of the lowest eigenvalue, compared with the value of the infrared cutoff (horizontal red line)
provided by the twisted mass. In the third strip we show the MC history
of the reweighting factor and its distribution.
One crucial parameter for stability issues and for controlling
discretization errors is the PCAC mass.
In the right plot of fig.~\ref{fig:algo_pcac}, we show the MC history and the distribution
of the PCAC mass at $x_0 = T/4$, together with the euclidean time dependence
of the PCAC mass.
It is remarkable that there is almost no sign of boundary O($a$) cutoff effects.
The analysis gives with the corresponding Z factors~\cite{ETMC_ren:2008in}
\be
%         A_1: &\quad&  am_{\rm PCAC} = 0.00073(22) \\
%         &\rightarrow& \quad M_{\rm R} = 0.0017(3) \\
am_{\rm PCAC} = 0.00045(12) \qquad \Rightarrow \qquad aM_{\rm R}^{\msbar}(2 {\rm GeV}) = 0.0012(2),
\ee
where
\be
  M_{\rm R}^{\msbar}(2 {\rm GeV}) = \frac{1}{Z_{\rm P}}M \qquad M = 
\sqrt{\left(Z_{\rm A}m_{\rm PCAC}\right)^2 + \mu_{\rm q}^2}.
\ee
%\vspace{-0.3cm}
We are clearly not at full twist. It is important
to remark that this is not so relevant in the
regime where chiral symmetry is restored.
Automatic O($a$) improvement~\cite{Frezzotti:2003ni} actually holds
in a finite volume and with suitable boundary conditions
also for massless Wilson fermions~\cite{Sint:2005qz}. This is somehow
related to the fact that in the region where chiral symmetry is restored
only O($am_{\rm PCAC}$) cutoff effects are expected, i.e. very small effects.
On the other hand if the mass is of O($a^2$) in general the latter could 
be the cutoff effects to become visible.
\subsection{Low energy constants}
\label{ssec:lec}
\begin{figure}[t]
\vspace{-0.8cm}
  \begin{center}
    {\hspace{-1.0cm}\includegraphics[height=2.0cm,angle=270,scale=4]{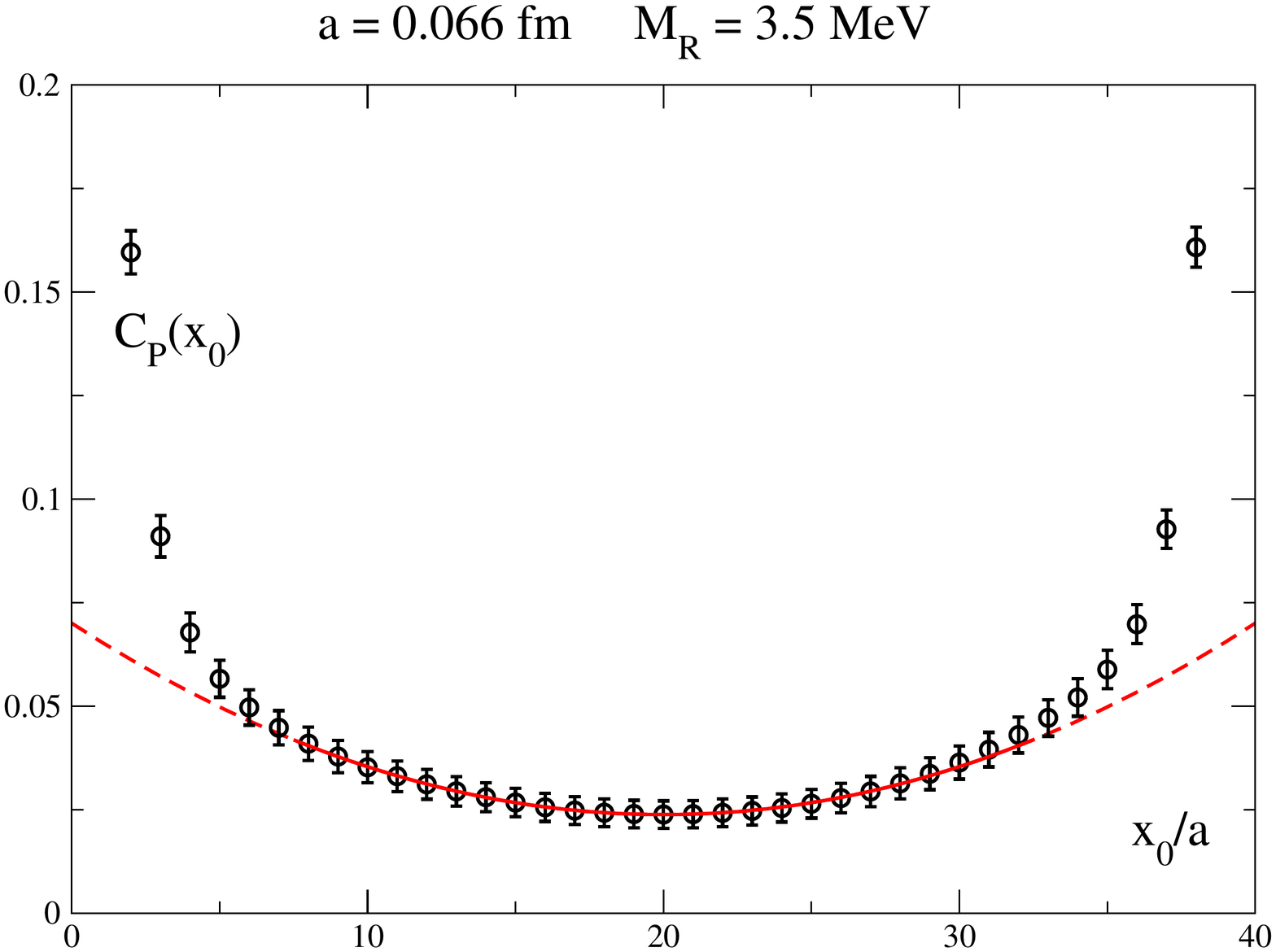}
    \includegraphics[height=2.0cm,angle=270,scale=4]{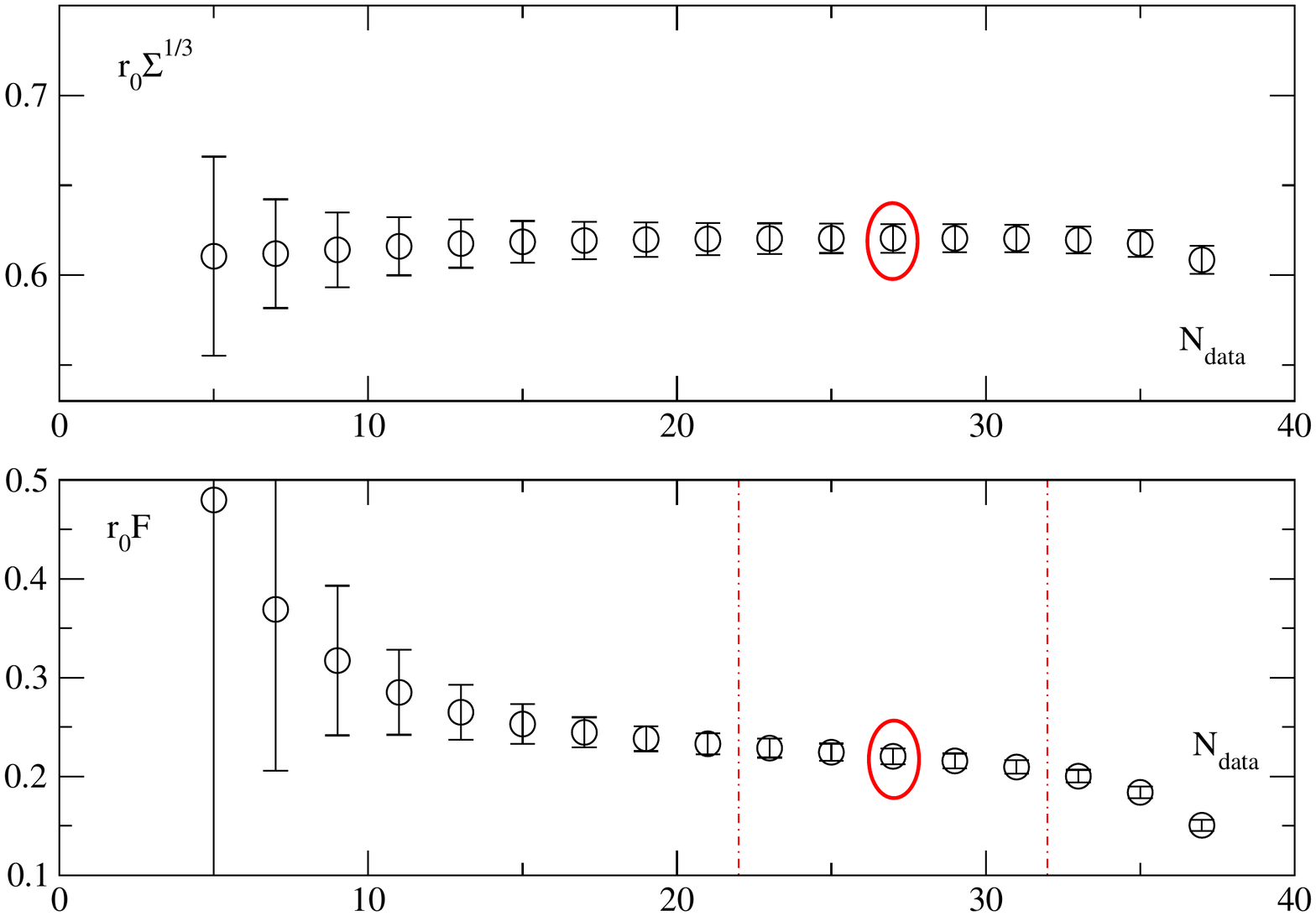}}
    \vspace{-0.8cm}
    \caption{{\it Left plot}: Euclidean time dependence of $C_{\rm P}(x_0)$ together
      with the result of the fit. The solid line indicates the fit range, while
      the dashed line indicates the same curve outside the fit range.
      {\it Right plot}: fit results for chiral condensate (first strip) 
      and decay constant (second strip) as a function of the number
      of points included in the fit around the middle point $T/2$. 
      The circles indicate the actual values quoted in the text and the dashed lines
    indicate the range of values used to determine the systematic error.}
    \label{fig:pp_fit}
  \end{center}
    \vspace{-0.5cm}
\end{figure}

The values of the low energy constants (LEC) can be extracted 
comparing the results of the numerical simulations 
for the euclidean time dependence of basic two-point functions
with the prediction of $\chi$PT~\cite{Hasenfratz:1989pk,Hansen:1990un}.
In this proceeding we consider the correlation function 
\be
  C_{\rm P}(x_0) = \frac{a^3}{L^3}\sum_{\bf x}C_{\rm P}(\bx,x_0) \qquad 
  \delta^{ab} C_{\rm P}(\bx,x_0) = \langle P^a(\bx,x_0)P^b({\bf 0},0)\rangle
\ee
between charged pseudoscalar currents
\be
  P^a(x) = \chibar(x)i\gamma_5\frac{\tau^a}{2}\chi(x).
\ee
The euclidean time dependence of the correlation function in $\chi$PT is given by
\be
  C_{\rm P}(x_0) = a_{\rm P} + \frac{T}{L^3}b_{\rm P}\left[\frac{y^2}{2}-\frac{1}{24}\right]+\ldots \qquad 
  y = \frac{x_0}{T}-\frac{1}{2},
\ee
where we have defined the following variables
\be
  a_{\rm P} = \frac{B_0^2F^4\rho^2}{8}G_1(u), \qquad b_{\rm P} = F^2B_0^2\left[1-\frac{1}{8}G_1(u)\right].
\ee
Details on the definitions of $\rho$, $u$ and $g_1$ can be found in~\cite{Hasenfratz:1989pk,Hansen:1990un}.
%and
%\be
%  u = 2B_0F^2MV\rho, \qquad \rho = 1 + \frac{3}{2}\frac{\beta_1}{F^2\sqrt{V}}, \qquad 
%  G_1(u) = \frac{8}{u}\frac{Y'(u)}{Y(u)}, \qquad Y(u) = \frac{2I_1(u)}{u}.
%\ee
We can thus fit the results from the numerical simulations with
the following fit formul\ae
\be
  C_{\rm P}(x_0) = A_0 + A_2y^2 \qquad \Rightarrow \qquad a_{\rm P} = A_0 + \frac{A_2}{12} \qquad 
  b_{\rm P} = A_2\frac{2L^3}{T}
\ee
In the left plot of fig.~\ref{fig:pp_fit}, we show the euclidean time dependence of $C_{\rm P}(x_0)$ together
with the result of the fit (red curve). 
The line is solid along the fit range and it becomes dashed 
where the points are not included in the fit.
The results of the analysis are
\be
  a^3L^3 A_0 = (5.94(36))\cdot10^{-3}, \qquad 
  a^3L^3 A_2 = (4.81(30))\cdot 10^{-2}
\ee
where the errors have been computed with a nested Jackknife/bootstrap procedure.
It is important to check the stability of the fit results
with respect to the number of data $N_{\rm data}$ included in the fit.
This is especially important if we want to make sure that the parabolic time
dependence is a real feature coming from simulating in the $\epsilon$ regime
and not just accidental, i.e. coming from the standard $cosh$ dependence
which can reproduce a fake parabolic behaviour around the $T/2$. 

In the right plot of fig.~\ref{fig:pp_fit}, we show the stability of the effective chiral condensate and
decay constant as a function of the number of data points (i.e time slices)
around the middle point included in the fits. 
While the chiral condensate shows a remarkably stable result
including more points in the fit, the decay constant
shows a somehow not completely flat dependence on the number of data
included in the fit.
Although this is not worrisome, it might be an indication 
of a physical volume not sufficiently large to suppress
higher order corrections.
A perfectly well defined way to proceed would be to include 
in the systematic error for $F$ the spread of 
its value in the region between the 2 dashed lines. 
%\begin{figure}[h]
%%\vspace{-1.0cm}
%  \begin{center}
%    {\includegraphics[height=2.0cm,angle=270,scale=7]{./r0sigmaF_405}}
%%    \vspace{-1.5cm}
%    \caption{Fit results for chiral condensate (first strip) 
%      and decay constant (second strip) as a function of the number
%      of points included in the fit around the middle point $T/2$. 
%      The circles indicate the actual values quoted in the text and the dashed lines
%    indicate the rangof values used to determine the systematic error.}
%    \label{fig:eps1}
%  \end{center}
%\end{figure}
The preliminary result of this analysis is
\be
  r_0\Sigma^{1/3} = 0.620(8), \qquad r_0F = 0.220(8)(10)
\ee
which agrees rather well with results
obtained in the $\epsilon$ regime using improved Wilson fermions~\cite{Hasenfratz:2008ce}.
%and with results obtained in the $p$ regime with Wtm~\cite{XXX}
%\be
%  r_0\Sigma^{1/3} = 0.617(15), \qquad r_0F = 0.224(10).
%\ee
%    \begin{table}[tb]
%      \centering
%      \begin{tabular}{|c|c|c|} \hline
%        Group  & $N_f$  &
%        $\Sigma$(2GeV)   \\ \hline
%        This work  & 2 & $-(282 \pm 4\; {\rm MeV})^3 $  \\
%        ETMC (2007) &  2  & $-(272 \pm 4 \pm 7 \; {\rm MeV})^3 $  \\
%        JLQCD (2007) &  2  & $-(251 \pm 7 \pm 11 \; {\rm MeV})^3 $  \\
%        Lang et al (2007) & 2 & $-(276 \pm 11 \pm 16 \; {\rm MeV})^3 $ \\
%        McNeile + MILC (2005) &  2+1  & $-(259 \pm 27  \; {\rm MeV})^3 $  \\
%        McNeile + JLQCD (2005) &  2   & $-(209 \pm 8  \; {\rm MeV})^3 $  \\
%       \hline
%      \end{tabular}
%    \end{table}

%    \begin{table}[tb]
%      \centering
%      \begin{tabular}{|c|c|c|} \hline
%        Group  & $N_f$  &
%        $F$   \\ \hline
%        This work  & 2 & $ 100(4)(5){\rm MeV} $  \\
%        ETMC(2007)  & 2 & $ 83(1)(3){\rm MeV} $  \\
%        QCDSF/UKQCD(2007)  & 2 & $ 79(5){\rm MeV} $  \\
%        JLQCD(2007)  & 2 & $ 78(3)(1){\rm MeV} $  \\
%        \hline
%      \end{tabular}
%    \end{table}
\section{Conclusions and outlook}
\label{sec:conclu}
We are establishing the basic knowledge to simulate
with Wilson-like fermions in the $\epsilon$ regime. To do this 
we have introduced a power counting to study the $\epsilon$ expansion
with Wilson-like fermions. The LO computation for the chiral
condensate confirms the absence of any phase transition,
and a NLO extension for the condensate and other 
observables is currently ongoing.

Numerical simulations in the $\epsilon$ regime with Wtm
are feasible using a PHMC with exact reweighting.
This particular choice allows to lower significantly the quark mass.
The extraction of LEC such as $\Sigma$ and $F$ becomes then feasible.
Moreover the NLO $\epsilon$ expansion is not contaminated
by chiral logs, which could be a benefit in reducing the systematic errors. 

Computations in this extreme region with Wilson-like fermions
require a detailed understanding of the
usual systematics: discretization errors, quark mass and volume dependence.
%We are planning to extend the analysis to other observables.

We remark that it might be advantageous to combine $p$ and $\epsilon$
regime simulations both as a tool to attack $2+1(+1)$ simulations, and
to further constrain the values of the LEC.

\acknowledgments

We thank the organizers of ``Lattice 2008'' for the very interesting
conference realized in Williamsburg. 
We thank J.~Gonzalez~Lopez, C.~Michael and G.C.~Rossi for a careful reading of this
proceeding, and all the members of ETMC for valuable discussions.

\bibliographystyle{JHEP-2}    % if you use h-elsevier.bst
\bibliography{lat_08}      % or whatever your .bbl file is

\end{document}